\documentclass[aps,amsmath,amssymb]{revtex4}
\usepackage{graphicx}
\usepackage{dcolumn}
\usepackage{bm}
\usepackage{psfrag}
\usepackage{graphicx}

\def\journal#1#2#3#4{{#1} {\bf #2}, #3 (#4)}
\newcommand{\be}{\begin{equation}}
\newcommand{\ee}{\end{equation}}
\newcommand{\bea}{\begin{eqnarray}}
\newcommand{\eea}{\end{eqnarray}}
\newcommand{\hf}{\frac12}
\newcommand{\nn}{\nonumber\\}
\def\eq#1{(\ref{#1})}
\def\la{\langle}
\def\ra{\rangle}

\def\Tr{{\mathrm{Tr}}}
\def\ord#1{{\cal O}\left(#1\right)}
\def\mr#1{{\mathrm{#1}}}
\def\v#1{{\bm{#1}}}
\def\dk{\Delta k}
\def\hphi{\hat\phi}

\def\hD{{\hat D}}
\def\ih{\frac{i}{\hbar}}
\def\sign{\mr{sign}}

\begin{document}
\title{Quantum renormalization group}

\author{S. Nagy$^1$, J. Polonyi$^2$, I. Steib$^1$}
\affiliation{$^1$Department of Theoretical Physics, University of Debrecen,
P.O. Box 5, H-4010 Debrecen, Hungary}
\affiliation{$^2$Strasbourg University, CNRS-IPHC, \\23 rue du Loess, BP28 67037 Strasbourg Cedex 2, France}
\date{\today}

\begin{abstract}
The observed IR and the spectator UV particles of a regulated, cutoff quantum field theory are entangled by their interactions; hence, the IR sector can be described by the help of the density matrix only. The tree-level renormalized trajectory is obtained for a self-interacting scalar field theory, containing the mixed state contributions. One needs a sharp cutoff in the momentum space as regulator to realize the true loss of information, caused by massive particles. 
\end{abstract}

\maketitle

\section{Introduction}
A many-body system can display an extremely rich dynamics which is covered partially by our observations, based usually on a few-particle subsystem only. Our goal is the reconstruction of the effective dynamics of the observed subsystem,  leaving the unobserved particles as spectators, simply an environment. This is aimed by the renormalization group (RG) method, the successive elimination of the unobserved, short scale fluctuations which was originally applied to eliminate the UV divergences of quantum field theories \cite{bogoliubov}. Such a strategy turned out to be very useful in statistical physics as well \cite{wilson} but it has been not realized that the mathematical setup of performing the blocking transformations on the partition function of classical statistical physical models where one integrates out recursively the variables within a block by keeping an average block variable fixed \cite{kadanoff}, is incomplete for the prescription of the  transition amplitude in quantum theories. What is missing in this scheme when applied to quantum systems is the mixed state contribution which arises due to the elimination of degrees of freedom  even if the full system is in a pure state \cite{ed}. 

The formalism of quantum field theory, based on the density matrix rather than transition amplitudes between pure states, is provided by Schwinger's closed time path (CTP) scheme \cite{schw} and it  has already widely been used to describe nonequilibrium physics \cite{keldysh,chou} and high energy physics \cite{calzetta}. This formalism, applied for a closed system, is well suited to calculate of the expectation values and its efficiency becomes evident when applied to open systems. 

When the high energy modes are not observed and they are eliminated then the remaining, observed degrees of freedom find themselves in an environment and their dynamics will be an open one. A modest step is presented in this paper in developing the RG scheme to find the mixed state contributions to the reduced density matrix of the observed system. The contributions of the eliminated degrees of freedom, calculated within the framework of the loop-expansion, usually start at the $\ord{\hbar}$, one-loop level, the only exception, known so far is the unstable, inhomogeneous mode in a vacuum with condensate \cite{vincent}. But there is actually a tree-level piece even without a condensate, only we need a bilocal action to capture it. This phenomenon is demonstrated here within the framework of the relativistic, scalar field theory, by the help of the functional RG method. The functional RG method can be realized in two different manners by using  sharp \cite{wh} or smooth cutoff \cite{wetterich,morris}. Most of the calculation, presented below, is performed with a sharp cutoff; however, a brief comparison with the smooth cutoff scheme is included, too.

Several publications have already been devoted to the study of the loop-corrections, generated by the generalization of the RG procedure to the CTP formalism. The generalization of the Wegner-Houghton equation, corresponding to a sharp cutoff where the evolution of the blocked action is followed in the local potential approximation was the subject of the earliest publications \cite{dalvit} which was restricted to the pure state contributions. The mixed state terms were kept in \cite{zanella} where the renormalized trajectory was considered for a scalar field theory, brought into contact with a heat bath. But the friction force has already been introduced in the  bare theory; hence, the mixed state contributions appear in a regular manner, without a gap. The master and the Langevin equations were derived for semiclassical gravity by coarse graining in Ref. \cite{calzettahu}. The RG strategy was applied in the CTP formalism and the one-loop blocking was used to derive the KPZ equation of stochastic quantization in Ref. \cite{zanellacalzetta}. The evolution generated by a sharp energy cutoff was derived for quantum dots \cite{gezzi} and in quantum mechanics \cite{aoki}. Such a cutoff does not correspond to the elimination of degrees of freedom and the mixed state contributions were generated by the heat bath only. 

The functional RG scheme \cite{wetterich,morris} is usually employed with smooth cutoff which keeps the state of the renormalized theory pure. Nevertheless some mixed state contributions can nevertheless be picked up by the coupling to an external heat bath. The exploration of nonthermal fixed points \cite{berges} and the loop-generated evolution of the effective action, kept in the local potential approximation \cite{mesterhazy} has already been achieved in this manner. Some spectral functions have been calculated in Ref. \cite{pawlowski} but the ansatz, the local potential approximation for the effective action, does not allow the mixed state contributions to emerge. The RG strategy can be used as a tool for a partial resummation of the perturbation series. When carried out in the CTP formalism may lead to a realization of the dynamical RG procedure \cite{boyanovsky}.

In Sec. \ref{ctps} we briefly recall the CTP formalism and its slight extension. The reader finds in Sec. \ref{blocks} the definition of the blocked action and its ansatz, used to defined the renormalized trajectory. The calculation of the tree-level effective action is reproduced in Secs. \ref{trees} and \ref{entls} contains the interpretation of the results. We turn briefly to the smooth cutoff scheme in Sec. \ref{smoothcfs}. Finally, the conclusions are given in Sec. \ref{concls}.

\section{Open quantum systems}\label{ctps}
The formalism, we need to handle open systems, can be reached in two steps, first the CTP scheme is worked out, followed by a slight generalization, the closed and open path (COTP) method, to deal with the reduced density matrix.

\subsection{CTP formalism}
One starts with a closed systems and uses the original CTP scheme \cite{schw}, developed for the calculation of the expectation values. It is based on the generating functional,
\be\label{genfunctcp}
Z[j^+,j^-]=\Tr[U(t_f,t_i;j^+)\rho_iU^\dagger(t_f,t_i;-j^-)],
\ee
where $\rho_i$ is the initial density matrix, $j^+$ and $j^-$ denote two sources, coupled linearly to the elementary variables, to generate expectation values and interaction vertices by functional derivation and $U$ stands for the unitary time evolution operator. The minus sign of $j^-$ is only a convention, motivated by simplicity. What makes this scheme particularly well suited for the calculation of the reduced density matrix is the unrestricted time evolution in contrast to the traditional formalism of quantum field theory where both the initial and the final states are fixed. The path integral expressions for the generating functional for a scalar field theory can be written in the form
\be\label{genf}
Z[\hat j]=\int D[\hphi]e^{\ih S[\hphi]+\ih\int dx\hat j_x\hphi_x},
\ee
where the CTP doublets, $\hat j=(j^+,j^-)$, $\hphi=(\phi^+,\phi^-)$ with the CTP ``scalar product'', $\hat j\hphi=\sum_\sigma j^\sigma\phi^\sigma$, have been introduced and the bare CTP action is given by $S[\hphi]=S_B[\phi^+]-S^*_B[\phi^-]$, $S_B[\phi]$ being the usual bare action. The integration is over closed time paths configurations, satisfying the condition $\phi^+_{t_f,\v{x}}=\phi^-_{t_f,\v{x}}$. The bare theory, regulated by an UV cutoff which assumes a large but finite value, $\Lambda$, is considered to be complete, namely its cutoff characterizes a complete set of physical states rather than our ignorance. As a result the bare theory is assumed to be closed and to have no environment. The convolution of $\hphi_{t_i,\v{x}}$ with the initial density matrix, $\rho_i$ and the details of the regularization of the path integral are suppressed. 

We take either the perturbative vacuum or a free, stationary thermal state as the initial state and the interactions are switched on adiabatically as $t_i\to-\infty$. The closing of the trajectories at the final time breaks the translation invariance in time and to recover this symmetry we go into the limit $t_f\to\infty$ where we can represent the trace at the final time in Eq. \eq{genfunctcp} by a redefinition of the free part of the CTP action, found by the help of the free propagator, 
\be\label{ctpfrprop}
i\hbar\hD_{0x,y}=\Tr\rho_i\begin{pmatrix}T[\phi_x\phi_y]&\phi_y\phi_x\cr\phi_x\phi_y&T[\phi_y\phi_x]^*\end{pmatrix},
\ee
defined by the free generating functional,
\be
Z_0[\hat j]=e^{-\frac{i}{2\hbar}\int dxdy\hat j_x\hD_{0x,y}\hat j_y}.
\ee
The identity 
\be
T[\phi_x\phi_y]+T[\phi_x\phi_y]^*=\phi_y\phi_x+\phi_x\phi_y,
\ee
imposes the condition $D^{++}+D^{--}=D^{+-}+D^{-+}$ which together with $D^{++}_{x,y}=-D^{--*}_{x,y}$ and $D^{+-}_{x,y}=-D^{-+*}_{y,x}$ yield the block structure,
\be\label{ctpprop}
\hD=\begin{pmatrix}D^n+iD^i&-D^f+iD^i\cr D^f+iD^i&-D^n+iD^i\end{pmatrix},
\ee
for the two-point function of any local operator in terms of three real functions, $D^n_{x,y}=D^n_{y,x}$, $D^f_{x,y}=-D^f_{y,x}$ and $D^i_{x,y}=D^i_{y,x}$, where $D^n$ and $D^f$ are called near and far Green functions by analogy with classical electrodynamics. One finds the Feynman propagator in the blocks $D^{++}=-D^{--*}$ and $D^{-+}=-D^{+-*}$ is given by the on shell Wightman function. These functions are the easiest to calculate in the operator representation where a straightforward calculation yields
\bea
D_{0p}^n&=&P\frac1{p^2-m^2},\nn
D_{0p}^f&=&-i\pi\delta(p^2-m^2)\sign(p^0),\nn
D_{0p}^i&=&-i(1+2n_p)\pi\delta(p^2-m^2),
\eea
for the Fourier transform,
\be
\hD_p=\int dxe^{ipx}\hD_{x,0},
\ee
where $P$ denotes the principal value prescription, $n_p=1/(e^{\beta\omega_p}-1)$ is the occupation number, and $\omega_p=\sqrt{m^2+\v{p}^2}$. 

The kernel of the free CTP action,
\be\label{freeact}
S_0[\hphi]=\hf\int dx dy\hphi_x\hat K_{x,y}\hphi_y,
\ee
is the inverse of the propagator, $\hat K=\hD^{-1}$ and the inversion, performed by the regulated Dirac-delta, $\delta_\epsilon(z)=\epsilon/\pi(z^2+\epsilon^2)$, yields the block structure,
\be\label{blockkern}
\hat K=\begin{pmatrix}K^n+iK^i&K^f-iK^i\cr-K^f-iK^i&-K^n+iK^i\end{pmatrix},
\ee
and the relations
\bea
D^{\stackrel{r}{a}}&=&(K^{\stackrel{r}{a}})^{-1},\nn
D^i&=&-D^rK^iD^a,
\eea
where $D^{\stackrel{r}{a}}=D^n\pm D^f$ stands for the retarded and advanced Greens functions and the notation $K^{\stackrel{r}{a}}=K^n\pm K^f$ is used for the kernel, too. We have 
\bea\label{ctpkernel}
K^n_p&=&p^2-m^2,\nn
K^f_p&=&i\epsilon\sign(p^0),\nn
K^i_p&=&i\epsilon(1+2n_p),
\eea
after ignoring $\ord{\epsilon^2}$ terms. The action of an interactive system, written as the sum of the free and the interaction pieces, $S[\phi]=S_0[\phi]+S_i[\phi]$, yields the CTP action,  $S[\hphi]=S_0[\hphi]+S_i[\phi^+]-S_i[\phi^-]$, for the path integral \eq{genf} where the integration is over independent trajectories, $\phi^\pm(x)$, with free initial and final conditions.

\subsection{COTP formalism}
In the second step the system, consisting of a particle of bare mass $m_s$, is brought into contact with its unobserved environment, made up by another kind of particle of bare mass $m_e$, and write action of the complete, closed system in the form $S[\phi,\chi]=S_{0s}[\phi]+S_{0e}[\chi]+S_{se}[\phi,\chi]$, where the fields $\phi$ and $\chi$ describe the system and the environment degrees of freedoms and $S_{0s}[\phi]$ and $S_{0e}[\chi]$ are the free action with $m^2=m^2_s$ and $m^2=m^2_e$, respectively. The reduced density matrix of the system, $\rho[\phi^+_f,\phi^-_f]$, is given by the partial trace over the environment,
\be\label{genfunctop}
\rho[\phi^+_f,\phi^-_f,\hat j]=\la\phi_f^+|\Tr_{env}[U(t_f,t_i;j^+)\rho_iU^\dagger(t_f,t_i;-j^-)]|\phi_f^-\ra.
\ee
It can be written as a path integral,
\be\label{genfrd}
\rho[\phi^+_f,\phi^-_f,\hat j]=\int D[\hphi]e^{\ih S_{eff}[\hphi]+\ih\int dx\hat j_x\hphi_x},
\ee
extending over open system trajectories, $\hphi_{t_f,\v{x}}=\hphi_{f\v{x}}$, and containing the COTP  effective action, 
\be\label{cotpeffact}
S_{eff}[\hphi]=\hf\int dx dy\hphi_x\hat K^d_{x,y}\hphi_y+S_i[\hphi].
\ee
The first term here is the free action, constructed by the help of the CTP diagonal kernel,
\be\label{kdiag}
\hat K^d_p=\begin{pmatrix}p^2-m^2+i\epsilon&0\cr0&-p^2+m^2+i\epsilon\end{pmatrix},
\ee
with $m^2=m_s^2$ and $S_i$ is defined by the CTP path integral,
\be
e^{\ih S_i[\hphi]}=\int D[\hat\chi]e^{\frac{i}{2\hbar}\int dx dy\hat\chi_x\hat K_{x,y}\hat\chi_y+\ih S_{se}[\phi^+,\chi^+]-\ih S_{se}[\phi^-,\chi^-]},
\ee
integrating over closed environment trajectories, $\chi^+_{t_f,\v{x}}=\chi^-_{t_f,\v{x}}$, and using the free kernel, \eq{ctpkernel}, with $m^2=m^2_e$. The unitarity of the time evolution for the full system assures that the predictions are independent of $t_f$ and the trace of the reduced density matrix remains unity.

\section{Blocking}\label{blocks}
The blocked theory is defined by a gliding sharp cutoff, $k$, reflecting our ignorance about or inability of observing the UV modes. The bare theory is supposed to be defined with the same cutoff at $k=\Lambda$. The bare action is assumed to be of the form
\be\label{bareact}
S_B[\phi]=\int dx\left[-\hf\phi_x\Box\phi_x-\frac{m_B^2}2\phi_x^2-U_B(\phi_x)\right],
\ee
where the bare potential, $U_B(\phi)$, is an even polynomial, starting beyond the quadratic order. The reduced density matrix of the blocked theory can be found by integrating out the unobserved modes and it defines the COTP effective theory according to \eq{genfrd}, where the integration extends over modes with spatial momentum $|\v{p}|<k$ and the blocked action is defined by Eqs. \eq{cotpeffact}--\eq{kdiag} with $m^2=m_B^2$ and the bare CTP path integral,
\be\label{bareeact}
e^{\ih S_i[\hphi]}=\int D[\hat\chi]e^{\frac{i}{2\hbar}\int dx dy\hat\chi_x\hat K_{x,y}\hat\chi_y-\int dx[U_B(\phi^+_x+\chi_x)-U_B(\phi^-_x+\chi_x)]},
\ee
with integration over the UV field, $\chi$, with spatial momentum $k<|\v{p}|$ and the free kernel $\hat K$, defined by Eqs. \eq{ctpkernel}, has $m^2=g_{B2}$. 

The CTP index appears like an internal quantum number in the Feynman rules and the lines of a graph represent either the diagonal or the off diagonal CTP blocks of the free propagator \eq{ctpfrprop}. To find the physical interpretation of a graph it is important to realize that the off diagonal CTP block of the propagator,
\be
\hD^{+-}_{x,y}=\sum_n\la n|\phi(x)\rho_i\phi(y)|n\ra,
\ee
represents the elementary processes where a particle is added or removed by $U$ and $U^\dagger$ in the generating functional of Eq. \eq{genfunctop}. Such a modification of the state appears in the distant future at the trace hence it must be on shell. Thus the lines which stand for the CTP off diagonal propagator and have UV spatial momentum represent the (de)excitations of the environment. 

It is instructive to distinguish three kinds of CTP Feynman graph \cite{effth}. The graphs without CTP off diagonal lines, $D^{\pm\mp}$, are called homogeneous. These graphs are identical to those of the traditional, transition amplitude based formalism. The graphs which have CTP off diagonal lines but all external legs belong exclusively either to $\phi^+$ or $\phi^-$ are called inhomogeneous. These graphs contribute such CTP Green functions which occur in the traditional formalism, too. Finally, a genuine CTP graphs has the CTP off diagonal lines and its external legs belong to both $\phi^+$ and $\phi^-$. If the initial state is the vacuum with a gap in the excitation spectrum then the energy conservation cancels the inhomogeneous graphs and the Green functions which are present in the traditional and the CTP formalisms are identical. Hence the new effective vertices of \eq{bareeact}, compared to the traditional formalism, arise either from the genuine CTP graphs including couplings between $\phi^+$ and $\phi^-$ and expressing IR-UV entanglement or from the heat bath. Both kinds of vertex represent entanglement and generate mixed state contributions.

The gliding cutoff poses a serious problem, namely we can not preserve the Lorentz symmetry in a theory with finite cutoff. The problem of an UV cutoff in the Minkowski space-time can be seen by noting that the volume of the space-time region which is bounded by two hyperboloids is infinite. Thus one can not define a region of finite volume by the help of invariants. Different regulators break the boost symmetry in different manner. For instance when the Wick rotation, following a regularization in the Euclidean space-time, is performed within a finite energy interval it generates an exponentially weak symmetry breaking which is acceptable in a renormalizable theory where $\Lambda\to\infty$ but remains problematic in an effective theory with a finite cutoff. There is a Lorentz invariant regulator which is based on higher order derivatives \cite{paulivillars} but the cancellation of the divergences in such a manner leads to physically unacceptable, negative norm states. Hence we assume a Lorentz noninvariant, sharp gliding cutoff, $|\v{p}|<k$, for the blocked theory and project the evolving blocked action after each blocking step onto a Lorentz invariant ansatz.

The blocked action is obtained in the RG strategy by subsequent infinitesimal blockings: We start with the bare action at $k=\Lambda$ and decrease the gliding cutoff in infinitesimal steps, $k\to k-\dk$. The infinitesimal blocking is therefore the transfer of the modes with momentum $k-dk<|\v{p}|<k$ from the system to the environment and the modification of the effective action is given by the equation
\be\label{blocking}
e^{\ih S_{k-\dk}[\hphi]}=\int D[\hat\chi]e^{\ih S_k[\hphi+\hat\chi]},
\ee
where $\phi$ and $\chi$ are nonvanishing for $|\v{p}|<k-\dk$ and $k-\dk<|\v{p}|<k$, respectively. This step yields a functional differential equation, a far too involved mathematical problem and one has to use an approximate equation, obtained by projecting it onto a functional space, defined by some ansatz. We shall use the form $S=S_1+S_2$, where $S_1$ is kept in the local potential approximation,
\be
S_1[\hphi]=\hf\int dxdy\hphi_x\hat K^d_{x,y}\hphi_y-\int dx[U(\phi^+_x)-U(\phi^-_x)],
\ee
involving the block-diagonal kernel in \eq{kdiag}, of the free action with $m^2$ being the running mass and a real, even polynomial potential starting beyond the quadratic order. The nonlocal part of the action can be organized within the cluster expansion. Since the nonlocality arises mainly due to the on shell modes we truncate this expansion at the bilocal cluster,
\be\label{bilans}
S_2[\hphi]=-\int dxdyV_{x-y}(\hphi_x,\hphi_y).
\ee

When a mode is transferred from the open to the close time path sector then we have to provide the  off diagonal blocks of the kernel in its free action which represent the trace operation in the generating functional\eq{genfunctop}. Therefore the kernel in the free action in the path integral on the right-hand side of Eq. \eq{blocking} is given by Eqs. \eq{ctpkernel}, using the running mass.

\section{Tree-level evolution}\label{trees}
To find the saddle point to the path integral \eq{blocking} we have to solve the equation of motion for $\hat\chi(x)$ for a given $\hphi(x)$. The scalar product of a function $f_x$ with a regular Fourier transform and $\chi^n_x$, the space-time integral of the product, is ${\cal O}(\dk^n)$ thus it is enough to look for the saddle points of the linearized equation of motion. We seek the solution of the linearized equation of motion, $\hD^{-1}\hat\chi=\hat L$, where $(D^{-1})^{\sigma,\sigma'}=(D_0^{-1})^{\sigma,\sigma'}-\delta^{\sigma,\sigma'}\sigma U''(\phi^\sigma)$ is the inverse of the environment propagator and
\be\label{source}
L_x^\sigma=\sigma U'(\phi^\sigma_x)-2\int dy\partial_{\phi_x^\sigma}V_{x-y}(\hphi_x,\hphi_y).
\ee
The substitution of the solution back into $S_k[\hphi+\hat\chi]$ yields $S_{k-\dk}=S_k+\Delta S_k$, with
\be
S_k-S_{k-\dk}=\frac{\dk}2\int dxdy\hat L_x(\hD^{(k)}P^{(k-\dk,k)}\hD^{(k)-1}P^{(k-\dk,k)}\hD^{(k)})_{x,y}\hat L_y,
\ee
where the projector,
\be
P^{(k_1,k_2)}_{x,y}=\int\frac{d^4p}{(2\pi)^4}\Theta(|\v{p}|-k_1)\Theta(k_2-|\v{p}|)e^{-i(x-y)p},
\ee
restricts the contributions to the environment momentum shell, $k-\dk<|\v{p}|<k$, and the environment propagator is
\be
\hD^{(k)}_{x-y}=\int\frac{d^4q}{(2\pi)^4}\delta(|\v{q}|-k)\hD_qe^{-i(x-y)q}.
\ee

\begin{figure}
\includegraphics[scale=.4]{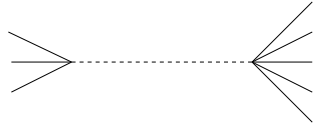}
\caption{An $\ord{\phi^{+3}\phi^{-5}}$ contribution to the right-hand side of Eq. \eq{trincr} which couples the time axes by the help of a fourth and and sixth order vertex. The two vertices are connected by a dashed line, standing for $D^{(k)+-}$.}\label{couplf}
\end{figure}

The projection onto our ansatz suppresses the $\phi$-dependence in $\hD^{(k)}$,
\be\label{trincr}
S_k-S_{k-\dk}=\frac{\dk}2\int dxdy\hat L_x\hD^{(k)}_{x-y}\hat L_y,
\ee
cf. Fig. \ref{couplf}, imposes the truncation
\be
L_x^\sigma\to\bar L_x^\sigma=\sigma U'(\phi^\sigma_x)-2\int dyW^\sigma_{x-y}(\hphi_y),
\ee
with
\be
W^\sigma_{x-y}(\hphi)=\partial_{\phi'^\sigma_x}V_{x-y}(\hphi',\hphi)_{|\phi'=0}
\ee
and yields the evolution equation
\be
\frac{d}{dk}S=\hf\int dxdy\hat{\bar L}_x\hD^{(k)}_{x-y}\hat{\bar L}_y.
\ee
The right-hand side of this equation can be written in form
\bea\label{eveqr}
\frac{d}{dk}S&=&\hf\sum_{\sigma,\sigma'}\int dxdy\biggl[\sigma U'(\phi^\sigma_x)D^{(k)\sigma,\sigma'}_{x-y}\sigma' U'(\phi^\sigma_y)-4\int dzW^\sigma_{z-x}(\hphi_x)D^{(k)\sigma,\sigma'}_{z-y}\sigma' U'(\phi^\sigma_y)\nn
&&+4\int dzdz'W^\sigma_{z-x}(\hphi_x)D^{(k)\sigma,\sigma'}_{z-z'}W^{\sigma'}_{z'-y}(\hphi_y)\biggr],
\eea
making the preservation of the two-cluster structure explicit.

To find the solution of the evolution equation we rewrite it first in momentum space,
\bea
\frac{d}{dk}S&=&\hf\int\frac{d^4q}{(2\pi)^4}\delta(|\v{q}|-k)\biggl[4[W^\sigma_{-q}(\hphi)]_{-q}D_q^{\sigma,\sigma'}[W^{\sigma'}_q(\hphi)]_q\nn
&&-4[W^\sigma_{-q}(\hphi)]_{-q}D_q^{\sigma,\sigma'}\sigma'[U'(\phi^\sigma)]_q+\sigma[U'(\phi^\sigma)]_{-q}D^{\sigma,\sigma'}_q\sigma'[U'(\phi^\sigma)]_q\biggr],
\eea
by the help of the Fourier transforms, $f_q=\int dxf_xe^{ixq}$. Note the simple structure of the evolution: for each value of the gliding cutoff the function $V_q$ is fixed within the Fourier space space region $k-\dk<|\v{q}|<k$ without using the other values of $V_q$. Thus the tree-level evolution simply accumulates the independent contributions recovered at each gliding cutoff value. The solution, corresponding to the initial condition \eq{bareact}, is $S_2$, given by Eq. \eq{bilans} with
\be\label{treebsc}
V_{x-y}(\hphi_x,\hphi_y)=-\hf\sum_{\sigma,\sigma'}\sigma\sigma'U'(\phi^\sigma_x)D^{(k,\Lambda)\sigma\sigma'}_{x-y}U'(\phi^{\sigma'}_y),
\ee
where
\be\label{htraj}
\hD^{(k,\Lambda)}_{x-y}=\int\frac{d^4q}{(2\pi)^4}\Theta(|\v{q}|-k)\Theta(\Lambda-|\v{q}|)\hD_qe^{-i(x-y)q}
\ee
is the propagator, covering all eliminated mode. The local potential does not evolve on the tree level, $U(\phi)=U_B(\phi)$ and $m^2=m_B^2$.

\section{Tree-level system-environment entanglement}\label{entls}
To find the physical interpretation of the tree-level contribution to the blocking we choose the vacuum as the initial state, place the system into a large but finite quantization box and write the running density matrix in the form
\be
\rho_k=\sum_np_n|\psi_n^{(0,k)}\ra\la\psi_n^{(0,k)}|
\ee
where the state $|\psi^{(k_1,k_2)}\ra$ belongs to the sector of the Fock-space which contain particles with spatial momentum $k_1<|\v{p}|<k_2$. The bare theory is closed thus the summation over $n$ is restricted to a single value for $k=\Lambda$ and $|\psi_1^{(0,\Lambda)}\ra=|0^{(0,\Lambda)}\ra$. Furthermore, we introduce a weak, external field, $\hphi$, and seek the leading order contribution to the running density matrix during the blocking, $k\to k-\dk$, by expanding in $\hphi$.

Let us first consider the change of the density matrix which is induced by the CTP diagonal contributions in \eq{treebsc} for a given $x$ and $y$. These terms describe the creation and annihilation of an UV particle at $x$ and $y$, a correction to $U$ and $U^\dagger$ in the generating functional \eq{genfunctop}, written in the form
\be\label{ddensm}
\Delta\rho_k=\begin{cases}0&t<t_1~\mr{or}~t_2<t,\cr
\sum_np_n\sum_{\hat m}|\psi^{(0,k-\dk)}_{n,m^+}\ra\otimes|\eta_{m^+}^{(k-\dk,k)}\ra\la\eta^{k-\dk,k)}_{m^-}|\otimes\la\psi^{(0,k-\dk)}_{n,m^-}|&t_1<t<t_2,\end{cases}
\ee
where $t_1=\min(x^0,y^0)$, $t_2=\max(x^0,y^0)$ and the sum is over the pairs, $\hat m=(m^+,m^-)$, $m^\pm\ge0$, excluding the case $\hat m=(0,0)$. The states $|\eta_m^{(k-\dk,k)}\ra$ with $m\ge1$ are normalized momentum eigenstates of an UV particle and $m=0$ denotes the UV vacuum state, $|\eta_0^{(k-\dk,k)}\ra=|0^{(k-\dk,k)}\ra$, hence $\la\eta^{k-\dk,k)}_{m^-}|\eta_{m^+}^{(k-\dk,k)}\ra=\delta_{m^+,m^-}$. The IR state, $|\psi^{(0,k-\dk)}_{n,m}\ra$, generated by the UV particle, is the relative state of $|\psi^{(0,k)}_n\ra$ with respect to $|\eta_m^{(k-\dk,k)}\ra$ \cite{everett}. A system-environment entanglement is generated for $t_1<t<t_2$ and it produces a mixed contribution to the state of the blocked theory,
\be
\Delta\rho_{k-\dk}=\begin{cases}0&t<t_1~\mr{or}~t_2<t,\cr
\sum_{m,n}p_n|\psi^{(0,k-\dk)}_{n,m}\ra\la\psi^{(0,k-\dk)}_{n,m}|&t_1<t<t_2.\end{cases}
\ee
Such an IR-UV entanglement and the resulting mixed states are properly given account in the traditional RG scheme because there are no environment excitations in the final state at $t_f>t_2$.

The off diagonal terms describe the creation of an UV particle in both $U$ and $U^\dagger$ with the same momentum and their sum yields
\be\label{oddensm}
\Delta\rho_k=\begin{cases}0&t<t_1,\cr
\sum_np_n\sum_{\hat m}|\psi^{(0,k-\dk)}_{n,m^+}\ra\otimes|\eta_{m^+}^{(k-\dk,k)}\ra\la\eta^{k-\dk,k)}_{m^-}|\otimes\la\psi^{(0,k-\dk)}_{n,m^-}|&t_1<t<t_2,\cr
\sum_np_n\sum_{\hat m_1,\hat m_2}|\psi^{(0,k-\dk)}_{n,m_1^+,m_2^+}\ra\otimes|\eta_{m_1^+,m_2^+}^{(k-\dk,k)}\ra\la\eta^{k-\dk,k)}_{m_1^-,m_2^-}|\otimes\la\psi^{(0,k-\dk)}_{n,m_1^-,m_2^-}|&t_2<t,\end{cases}
\ee
where $|\eta_{m_1,m_2}^{(k-\dk,k)}\ra$ denotes a two UV particle state for $m_1,m_2\ne0$, a single UV particle state if one of the indices is 0 and the UV vacuum when both indices are 0. Now the system-environment entanglement and the mixed terms in the reduced density matrix,
\be\label{reddm}
\Delta\rho_{k-\dk}=\begin{cases}0&t<t_1,\cr
\sum_{m,n}p_n|\psi^{(0,k-\dk)}_{n,m^+}\ra\la\psi^{(0,k-\dk)}_{n,m^-}|&t_1<t<t_2,\cr
\sum_{m_1,m_2,n}p_n|\psi^{(0,k-\dk)}_{n,m_1,m_2}\ra\la\psi^{(0,k-\dk)}_{n,m_1,m_2}|&t_2<t,\end{cases}
\ee
persist until the final time and indicate that the environment is left in an excited state. These contributions require the COTP formalism.

The CTP diagonal terms to the blocking are complex and their real and imaginary components are proportional to $D^n_{x-y}$ and $D^i_{x-y}$, respectively, the latter representing the finite lifetime of the environment excitations. Such a contribution would be interpreted in the traditional formalism of quantum field theory as the violation of the unitarity of the time evolution. But the CTP off diagonal contributions are imaginary, as well, and restore the unitarity of the time evolution for the reduced density matrix. 

The excitations at the final time, $t_f\to\infty$ must preserve the energy and are on shell. The distinguished feature of the on shell excitations is that they can be captured as $\ord{\hbar^0}$ saddle point contributions, cf. Fig. \ref{couplf}. It is remarkable that the hallmark of quantum physics, the system-environment entanglement, is generated on the tree-level. This contribution is described by classical field theory, by the radiation field, generated by the far field Green function, $D^f$, in the language of classical electrodynamics.

\section{Smooth cutoff}\label{smoothcfs}
The functional RG method with smooth cutoff is an algorithm to solve quantum field theoretical models by interpolating along an arbitrary, nonphysical, one parameter family of systems between a soluble model and our bare theory. This strategy can be realized by the help of either the bare \cite{polch} or the effective action \cite{wetterich,morris}. We extend the latter scheme to allow unrestricted final state and the evolution of the effective action is followed as the modes are turned on gradually from a hardly fluctuating, perturbative model until we reach the physical theory. The final result is independent of the path used in the process as long as the evolution equation is solved exactly. In other words, the modes are eliminated with a freely chosen dispersion relation but the effects of this arbitrary feature cancel between the subsequent steps of the evolution. 

The interpolating theories are constructed by inserting a suppressing quadratic term into the bare action, $\hD^{-1}_0\to\hD^{-1}_0+\hat R_k$ with $\hat R=\mr{Diag}(R_k,-R_k)$, $k$ parameterizing the interpolation. In order to recover some qualitative similarity with the scale dependence, generated by a sharp cutoff, one requires that the Fourier transform of the suppression, $R_{k,q}$, is large for $|\v{q}|\ll k$ and is small when $|\v{q}|\gg k$. The former makes the field weakly fluctuating and renders the theory soluble for large $k$ and the latter assures that the original theory is recovered as $k\to0$. The system remains closed during the evolution of $k$ from its large initial value to 0 and the $k$-dependence of its expectation values is generated by the $k$-dependent dispersion relation of its free particles. One follows in this scheme the effective action,
\be\label{effact}
\Gamma[\hphi]=W[\hat j]-\int dx\hphi_x\hat j_x,
\ee
defined by the help of $W[\hat j]=-i\ln Z[\hat j]$, given by the generating functional, \eq{genf}, and $\hphi=\delta W /\delta\hat j$. 

The path integral \eq{genf} possesses a saddle point but we do not have a small parameter to organize the nonlinear terms in its equation of motion. In fact, such a small parameter, $\dk/k$, owes its existence in the sharp cutoff scheme to the complete freeze out of the modes $|\v{p}|<k-\dk$ during the blocking. By assuming that the saddle point has a small amplitude, $\hat\chi\sim0$, one can easily calculate the effective action in $\ord{\hat\chi^2}$ by following the steps, outlined in  the case of a sharp cutoff. The result for $\Gamma[\hphi]$ is the expression \eq{treebsc}, except the replacement $\hD^{(k,\Lambda)}\to\hD^{R_k,\Lambda}$, where
\be\label{htrajs}
\hD^{R_k,\Lambda}_{x-y}=\int\frac{d^4p}{(2\pi)^4}\Theta(\Lambda-|\v{p}|)\frac{e^{-i(x-y)p}}{\hD^{-1}_p+\hat R_{k,p}}.
\ee
The local potential receives no contributions similarly to the case of the sharp cutoff hence $U(\phi)=U_B(\phi)$ and $m^2=m_B^2$. 

The sharp and the smooth cutoff schemes are complementary in the following sense. A disadvantage of the sharp cutoff scheme is the impossibility of going beyond the local potential approximation and including higher order contributions of the gradient expansion in the blocked action. Furthermore, the blocked action, used in the path integral, offers little help in finding the expectation value of observables. Another feature of this scheme which somehow restricts its application is that the genuine CTP contributions are on shell and describe the excitations in the asymptotic state of the environment which take place beyond the cutoff. To generate such an excitation we need a large number of IR excitations if the cutoff is high. The gradient expansion is not limited, important expectation values, the one particle irreducible vertex functions, are known and arbitrarily small energy excitations are eliminated in the smooth cutoff scheme. However this scheme has two important shortcomings. It follows interpolating theories, an arbitrary family of models, and only the end point of the evolution, $k=0$, belongs to the physical case. Furthermore, it is impossible to separate the states to be eliminated and retained in the Fock space. This prevents us from looking for the system-environment entanglement, from defining the reduced system density matrix and from finding the mixed system states. Nevertheless the density matrix can be calculated and the expressions are formally similar to Eqs. \eq{ddensm} and \eq{oddensm}, except that the system state remains pure. We follow here the interpolation of the density matrix of a closed system as the dispersion relation of the particles is gradually changed.

\section{Conclusions}\label{concls}
We believe that the RG procedure, embedded into the CTP formalism opens up several issues which appeared to be complete and closed in the traditional renormalization in quantum field theories. The reason is that the renormalized theory corresponds to an open system, the monitored, low energy particles interact with the unobserved high energy sector. The IR-UV entanglement, generated in such a manner, produces a mixed state for the IR sector. Although it is always possible to reconstruct the expectation values by the help of transition amplitudes this procedure is rather cumbersome in practice. In the light that the ultimate theory of everything is unknown and we are always dealing with effective field theories theories in physics the IR-UV entanglement seems to be a generic feature of our observations. As opposed to the traditional RG method the CTP improved renormalization scheme can give account of the system-environment entanglement which has no counterpart in classical statistical physics; therefore, we call the procedure, presented above, the quantum renormalization group.

The IR-UV entanglement, addressed here by following the cutoff-dependence of the bare action within the framework of the functional RG method, equipped with sharp gliding cutoff, originates from processes, taking place above the cutoff and this feature limits its importance. For instance an $\ord{\phi^n}$ term in the action generates IR-UV entanglement if the energy of $n-1$ IR particles is sufficient to create an UV particle, a weak effect for high cutoff. But the IR-UV entanglement takes place in theories with nonpolynomial action, such as the nonlinear sigma model and gauge theories with compact gauge group, as soon as the energy of a particle at the sharp cutoff scale is available in a finite space region. A natural application of the calculation of the blocked, bare action by taking into account the IR-UV entanglement are the high energy threshold effects, such as relativistic corrections.

Entanglement effects persist for long time if the environment is soft, ie. it supports low energy excitations and we may encounter irreversibility and dissipation when the environment is gapless. Such a situation can better be handled by calculating the effective action for all environment modes. This scheme motivates the use of smooth cutoff where the truncation of the functional evolution equation is less problematic.

We mention two lessons to be learned from the calculation, presented in this work. One point is the importance of facing the loss of information, resulting from the unobserved high energy modes. This generates entanglement and mixed state components which have observable effects but can be recovered by a cutoff which is constructed by the careful modelization of the effects of the unobserved particles in the measuring device. In the case of an environment with a gap in its excitation spectrum one needs a sharp cutoff in an obvious manner. Another, more technical point concerns the emergence of the tree-level contributions to the blocking relations. Such contributions to both the pure and the mixed state components appear with a gap and are missed in the gradient expansion, employed in the usual RG calculations. 

Finally, the basic, open question: What happens when usual one-loop contributions are retained? An open system displays a richer dynamics and has more coupling constants than a closed one. Are there new fixed points, universality classes, by taking into account the mixed state components?

\section*{ACKNOWLEDGMENTS}
S. Nagy acknowledges financial support from a J\'anos Bolyai Grant of the Hungarian Academy of Sciences, the Hungarian National Research Fund OTKA (K112233).

\end{document}